\documentclass[final,1p,times]{elsarticle} 

\usepackage{graphicx}
\usepackage{amssymb} 
\usepackage{amsthm} 
\usepackage{lineno}
\usepackage{color}
\usepackage{soul}

\journal{Nuclear Physics A} 

\begin{document}

\begin{frontmatter} 

\title{Dilepton production in high energy heavy ion collisions with 3+1D relativistic viscous hydrodynamics}
\author{G. Vujanovic{$^1$}, C. Young{$^1$}, B. Schenke{$^2$}, S. Jeon{$^1$}, R. Rapp{$^3$}, and C. Gale{$^1$}}
\address{{$^1$} Department of Physics, McGill University, 3600 University Street, Montreal, Quebec H3A 2T8, Canada\\
{$^2$}Physics Department, Brookhaven National Laboratory, Upton, NY 11973, USA\\
{$^3$}Cyclotron Institute and Dept. of Physics \& Astronomy, Texas A\&M University, College Station, TX 77843-3366, USA}

\date{\today}

\begin{abstract}
We present a first calculation of the dilepton yield and elliptic flow done with 3+1D viscous hydrodynamical simulations of relativistic heavy ion collisions at the top RHIC energy. A comparison with recent experimental data from the STAR collaboration is made. 
\end{abstract}

\end{frontmatter} 
\section{Introduction}
The lepton pairs resulting from nuclear collisions at energies typical of the Relativistic Heavy Ion Collider (RHIC) and of the Large Hadron Collider (LHC) will come from various sources which include high-temperature QCD processes and  reactions involving thermal hadrons in the confined sector. The decay of open-charm mesons  constitutes an irreducible background which  must be included in theoretical analyses, in the absence of direct vertex detection. 
The analyses now available from RHIC and soon to be done at the LHC will provide a wealth of observables to qualify the enduring success of relativistic hydrodynamics, and to verify whether  this successful phenomenology extends to electromagnetic observables. We present the results of a study of  dilepton production from a hot environment with finite shear viscosity. 

\section{Thermal dilepton production rates and their viscous correction}
\label{rates}

In this work, the quark-antiquark annihilation at leading order (the Born approximation) is used as the dilepton production rate at high temperatures. To include shear viscosity, the thermal distribution functions $n$ are augmented to include a correction $\delta n(p) = C \frac{\eta}{s}\frac{1}{2T^3}n(p)(1 \pm n(p))p^\alpha p^\beta \frac{\pi_{\alpha \beta}}{\eta}$ \cite{dusling-lin}. The rates now depend on $\pi^{\mu\nu}$, the viscous correction to the stress-energy tensor, and therefore  on details of the dynamical viscous hydrodynamical simulations. We use \textsc{music},  the three-dimensional hydrodynamical simulation of Ref. \cite{schenke-jeon-gale}. 
In the phase where composite hadrons exist and interact, the dilepton thermal production rate is related to the in-medium behaviour of the current-current correlator \cite{kapusta-gale-book}, the Wightman function. Using the Kubo-Martin-Schwinger (KMS) relation, one can write the lepton pair production rate, $R$,  in terms of the imaginary part of the retarded photon self-energy, ${\rm Im} \Pi^{\rm (R)}_{\mu \nu}$:
\begin{eqnarray}
\frac{d^4R}{d^4 q}= - \frac{\alpha}{12 \pi^4}\frac{1}{M^2} {\rm Im\, } \Pi^{{\rm (R)}\, \mu}_\mu \frac{1}{e^{\beta q^0}-1} \label{eq:HG_rate}
\label{idealHGRate}
\end{eqnarray}
where the lepton rest masses have been set to zero,   $\alpha$ is the electromagnetic fine structure constant,  and $q^2 = M^2, $ $M$ being the dilepton invariant mass. An equation involving the imaginary part of the retarded in-medium vector propagator, ${\rm Im}\,D_V^{\rm R}$, is obtained from Eq. (\ref{idealHGRate}) using the Vector Dominance Model (VDM). Here, we will not extend the definition of the Wightman function nor that of the KMS relation in the presence of viscosity. Instead we explore the changes viscosity entails by modifying the thermal self-energy in the propagator and the Bose distribution in Eq.~(\ref{idealHGRate}), both via $\delta n$. The total self-energy is $\Pi^{\rm tot}_{V} = \Pi^{\rm vac}_V +  \Pi^{\rm T}_V + \delta\Pi^{\rm T}_V$. The calculations of $\Pi^{\rm vac}_V$ are presented in Ref. \cite{eletsky-belkacem-ellis-kapusta,martell-ellis,vujanovic-ruppert-gale}; there, effective Lagrangians describe all interactions contributing to $\Pi^{\rm vac}_V$. For the in-medium self-energy $\Pi^T_V$,  we are using here the approach in Refs. \cite{eletsky-belkacem-ellis-kapusta, jeon-ellis}, where a forward scattering amplitude $f_{Va}(s)$ is used to obtain the thermal self-energy of the vector meson $V$. The self-energies of Ref. \cite{Rapp:1999us}  will  be considered in upcoming work. Finally,  $\delta \Pi^T_V$ is the correction to the  self-energies that stems from the inclusion of viscous effects.

\section{Dilepton production from charm decays}
\label{charm}

In proton-proton and Au+Au collisions with center-of-mass energies of 200 GeV, dileptons produced from the semi-leptonic decay of pairs of charm quarks dominate the yield in the ``intermediate mass range'' (between the $\phi$ and the $J/\psi$).  We use \textsc{pythia8} to generate events with heavy quarks. 
Note that \textsc{pythia} includes the important processes of flavour excitation as well as the radiative splittings important for large $p_T$. We also use \textsc{eks98} to determine the initial parton distribution functions in the nuclei. Then, using the same hydrodynamical description as was used to determine the thermal dilepton production, the heavy quarks are evolved using relativistic Langevin dynamics and the heavy quark spatial diffusion coefficient $D_c = 3/(2\pi T)$. The heavy quarks then hadronize. The procedure is given in more details in Ref. \cite{Young:2011ug}. 

\section{Results}
\label{production}

\begin{figure}[!h]
\vspace{0.50cm}
\begin{center}
\begin{tabular}{c c}
\includegraphics[scale=0.23]{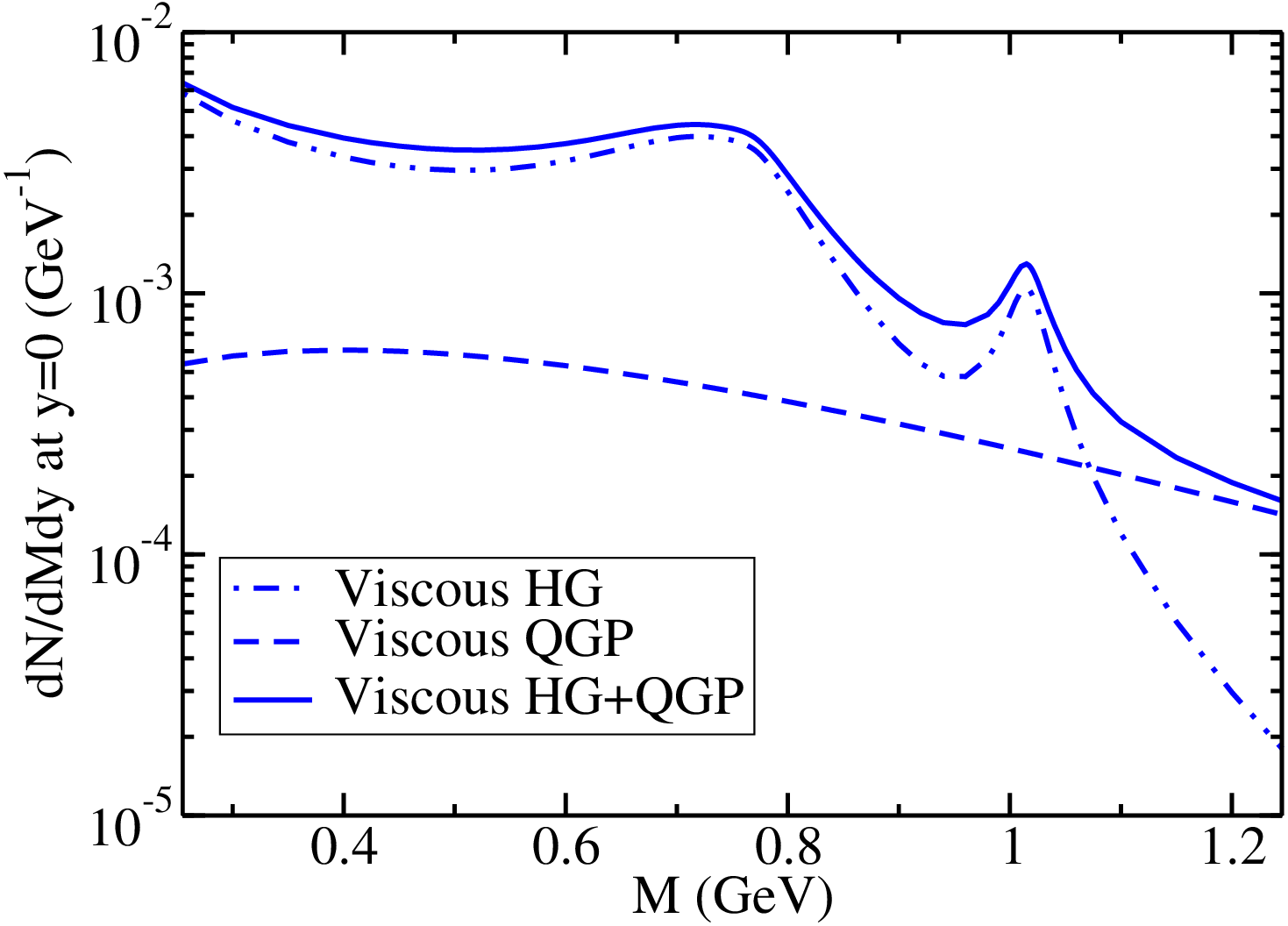} & \includegraphics[scale=0.23]{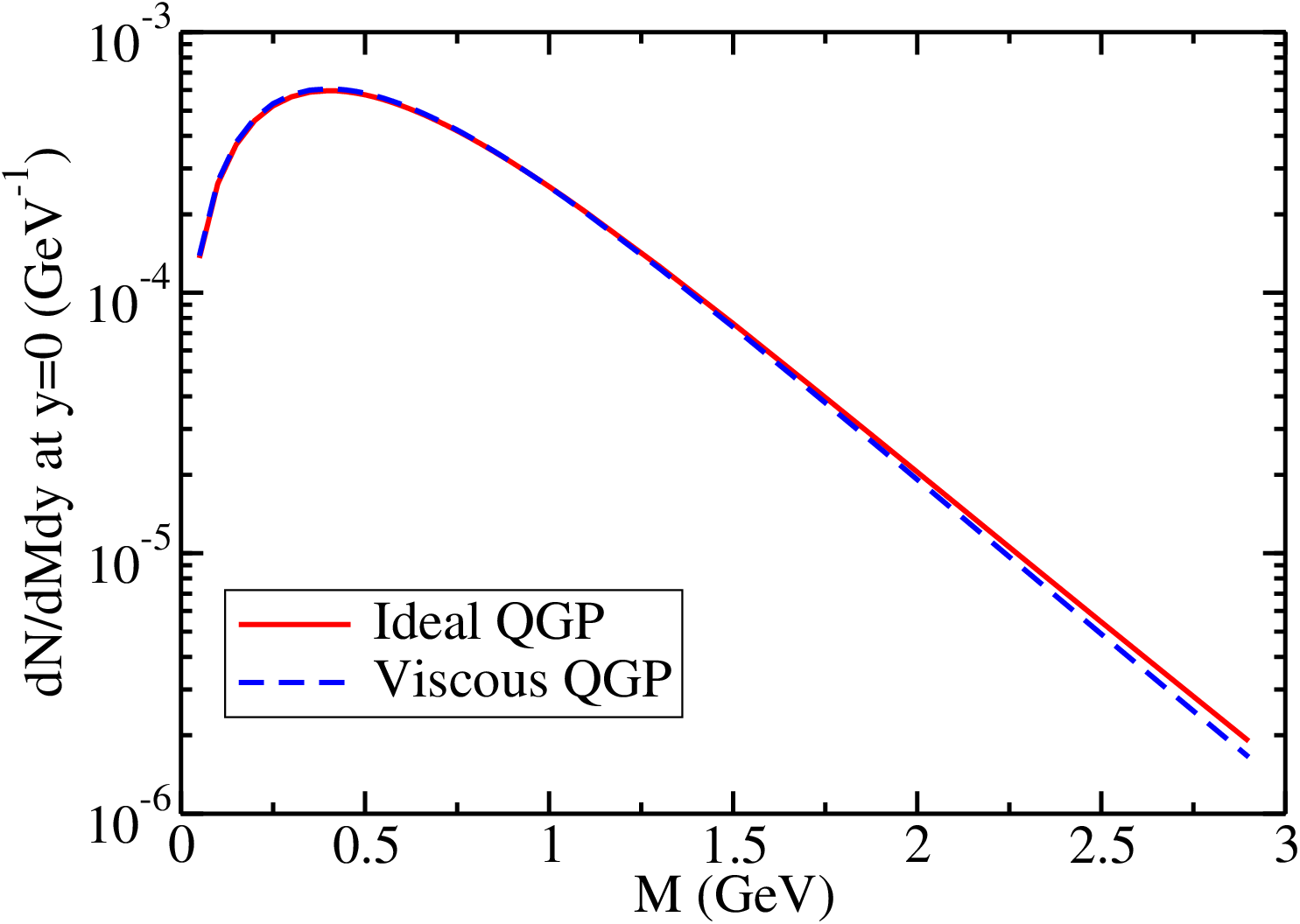}
\end{tabular}
\end{center}
\vspace{-0.75cm}
\caption{(a) Dilepton yield from the the hadronic and QGP phases  as a function of invariant mass. (b) Dilepton yield from the QGP extending to high invariant masses: ideal (solid curve) and viscous (dashed curve) are presented. All results shown here are for a minimum bias centrality class. }\label{pic:yield_ideal_vs_visc_M_hg_qgp}
\end{figure} 
 
 We show results with viscous and inviscid hydrodynamics, with $\tau_0$ = 0.4 fm/c, and consider Au + Au collisions at the top RHIC energy: $\sqrt{s}_{\rm NN}$ = 200 GeV. In the viscous simulations, the generation of entropy will impose a lower initial temperature in order to reproduce the final state hadronic observables. The ratio of shear viscosity to entropy density is set to $\eta/s = 1/ 4 \pi$. 
 
Figure \ref{pic:yield_ideal_vs_visc_M_hg_qgp} (a) presents the yield of dielectrons from the HG and QGP phases, with viscous effects. For the viscous HG calculation, we verified that the viscous corrections have little to no effect on the dilepton yield by correcting the numerator and denominator of Eq. (\ref{idealHGRate}) together, and individually. This result is in line with that obtained for real photons \cite{Dion:2011pp}, and it is consistent with a small $\pi_{\mu \nu}$ in the HG phase.   In the QGP phase, where $\pi^{\mu\nu}$ is more significant, the shape should be changed in the region where the viscous correction to the thermal rate is large (at large invariant masses). This can be seen for $M\sim2-3$ GeV in Fig. \ref{pic:yield_ideal_vs_visc_M_hg_qgp} (b). However, shear viscosity effects on dilepton yields remain modest.

\begin{figure}[!h]
\vspace{0.5cm}
\begin{center}
\begin{tabular}{c c}
\includegraphics[scale=0.23]{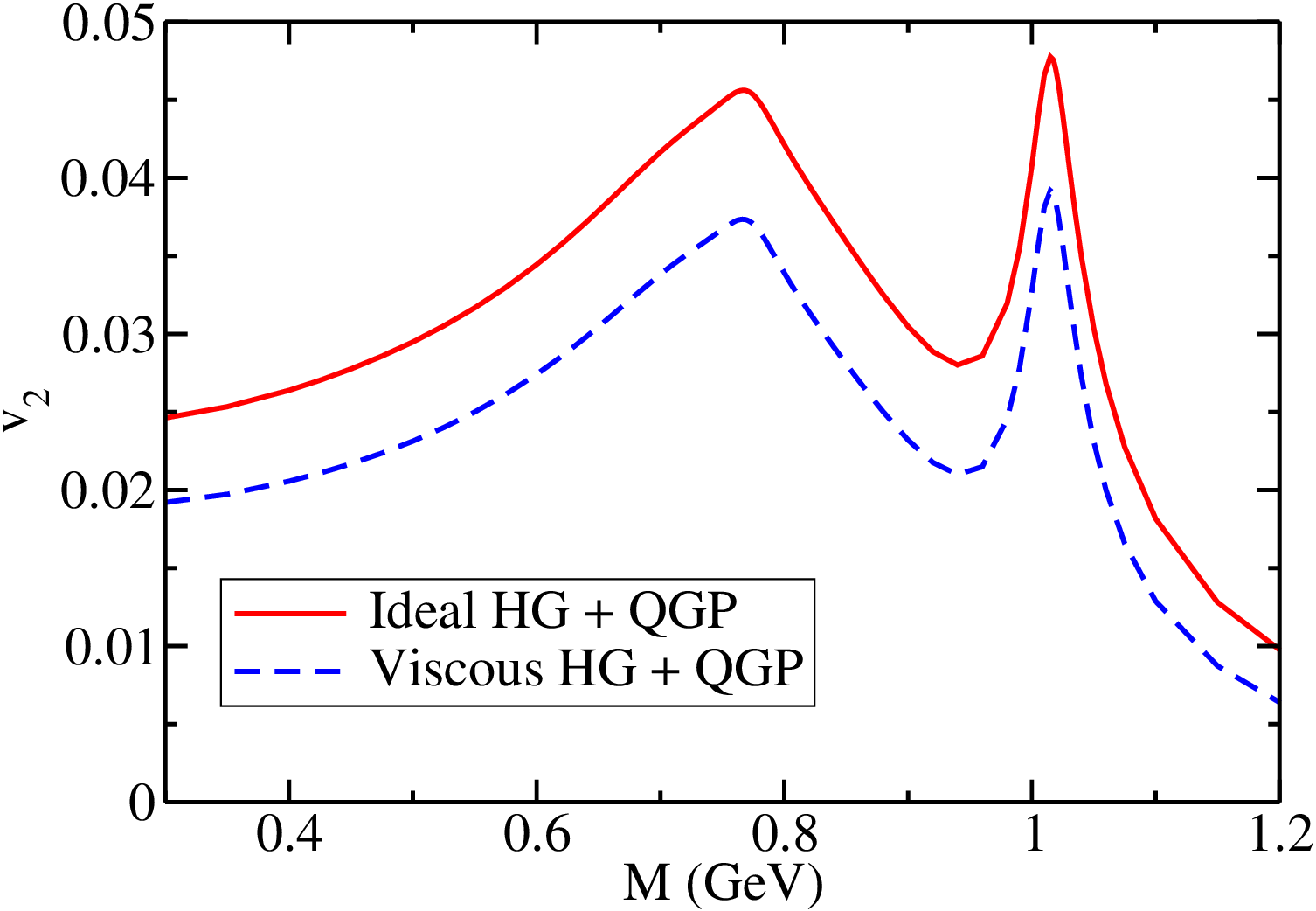} \includegraphics[scale=0.23]{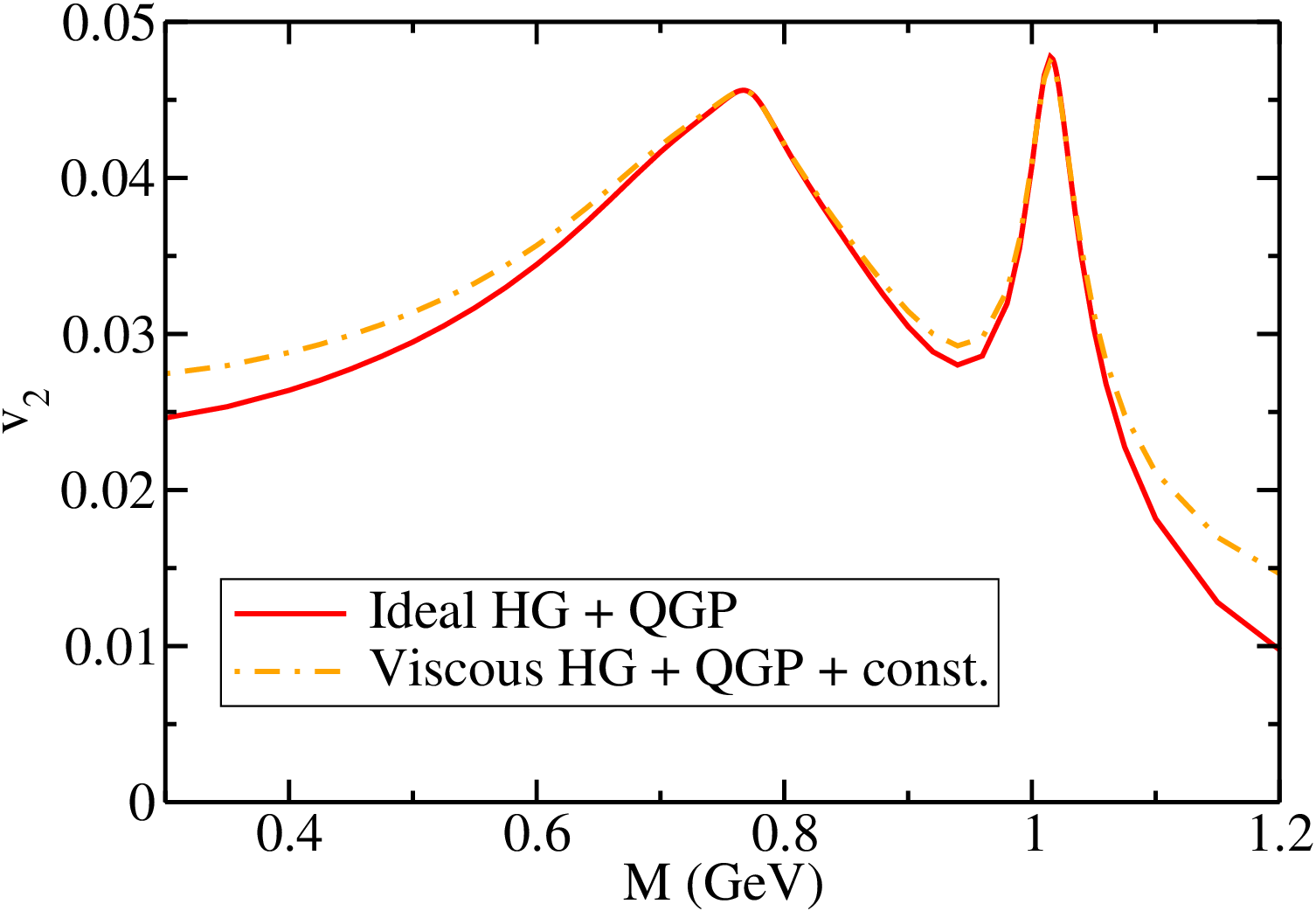}\\
\end{tabular}
\end{center}
\vspace{-0.75cm}
\caption{(a) Comparison of ideal (solid curve) and viscous (dashed curve) $v_2$ of dileptons. (b) Same as in (a), but with viscous results shifted by a numerical constant.  All results shown here are for a minimum bias centrality class. }\label{pic:v2_ideal_rapp_visc}
\end{figure}

The elliptic flow of dileptons is more sensitive to viscous effects than their spectra, as was also found for real photons \cite{Dion:2011pp}. Figure \ref{pic:v2_ideal_rapp_visc} shows that the peaks around the $\rho$, $\omega$,and $\phi$ masses first seen in a calculation using vacuum spectral distributions \cite{chatterjee-srivastava-heinz-gale} are present in $v_2(M)$, but are significantly reduced owing to in medium effects. The shear viscosity effects on the dynamics  slightly broadens the $M$ distribution (Fig. 2 (b)), as the temperature drop there is slower \cite{Dion:2011pp}. Note finally here that our $v_2 (M)$ calculation is approximate, in the absence of a formal extension of the KMS relation to out-of-equilibrium environments. 

\section{Comparison with STAR data}
\begin{figure}[!h]
\vspace{0.5cm}
\begin{center}
\begin{tabular}{c c}
\includegraphics[scale=0.23]{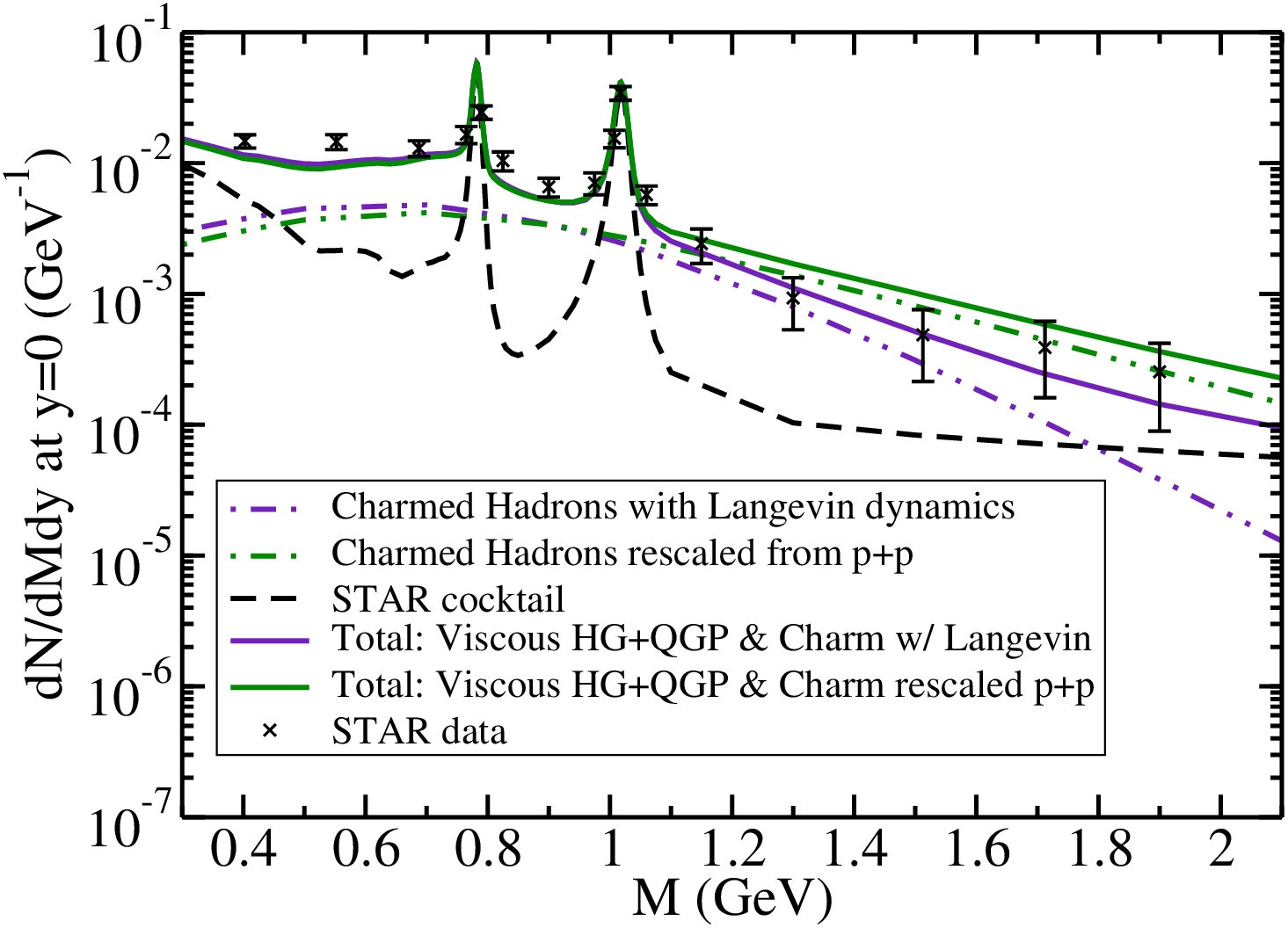} \hspace{0.3cm} \includegraphics[scale=0.23]{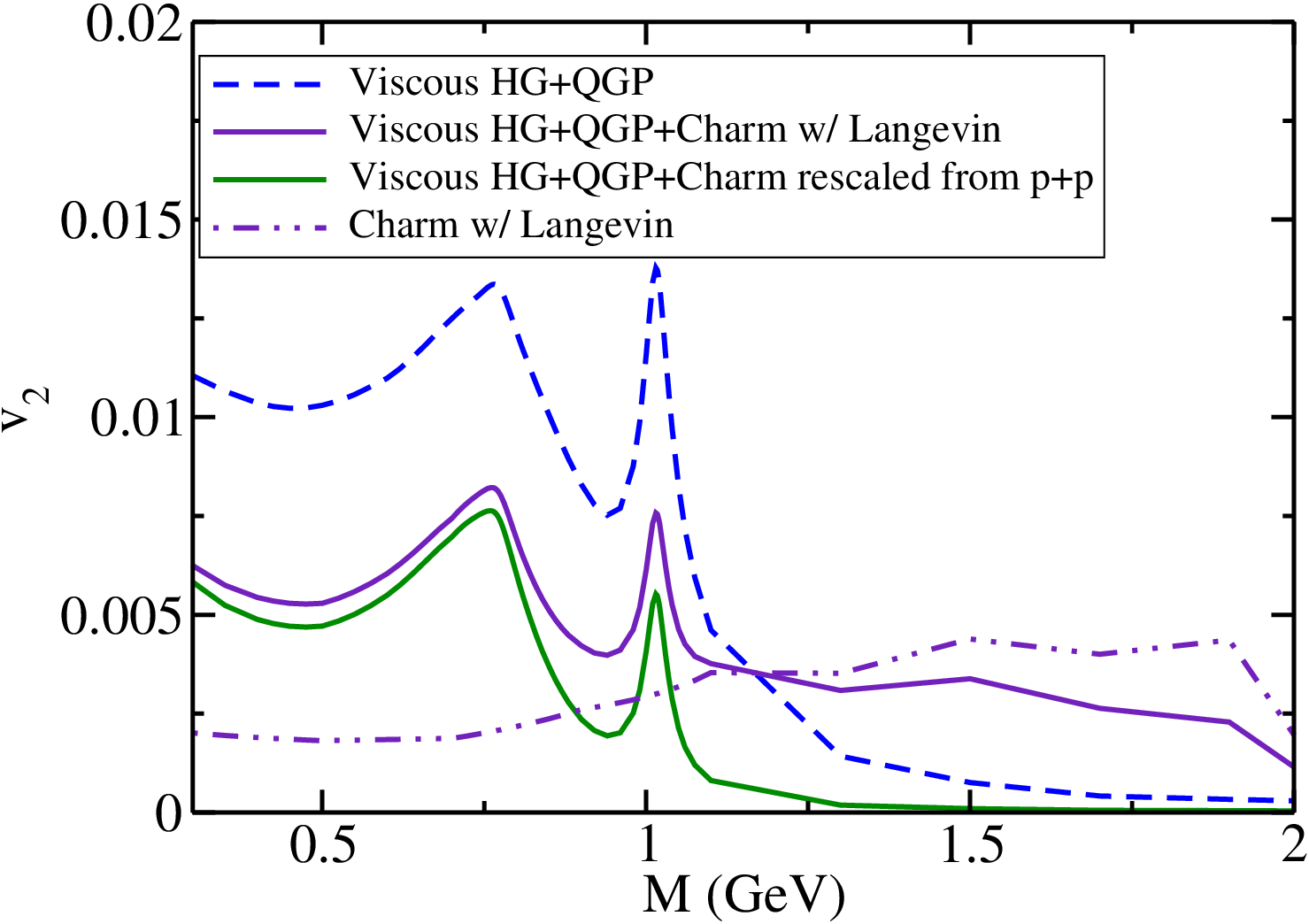}\\
\end{tabular}
\end{center}
\vspace{-0.75cm}
\caption{(a) Comparison of the dilepton yield as a function of $M$ with experimental data from STAR \cite{Dong}. (b) Dilepton $v_2$ versus invariant mass including charmed hadrons. All data and calculations shown here are for a 0-10\% centrality class. }\label{pic:yield_v2_charm}
\end{figure}   

The plots in Figure \ref{pic:yield_v2_charm} show the dielectron yield and $v_2$ as functions of invariant mass, now including the contribution from charm decays. Heavy quark energy loss has 
a significant effect on the dilepton yields in the intermediate mass range: the invariant mass spectrum is reduced by almost an order of magnitude. In addition, the flow of heavy quarks leads to an additional azimuthal anisotropy of the lepton pairs. Including the effect of energy loss leads to a better agreement with the current
 data on $dN/dM dy$ for intermediate mass dileptons from STAR \cite{Dong}. It is fair to write, however, that both scenarios are still consistent with the data, given the current experimental error bars.  Finally, not included in this first analysis are the possible effect of higher mass thermal hadrons on the yield and on the elliptic flow of intermediate mass dileptons \cite{Li:1998xn}. 

\section{Conclusion}
We have presented a calculation of lepton pair production made using a 3+1D relativistic hydrodynamics approach, with finite shear viscosity. We have  shown the effect of charmed quark energy loss on dilepton spectra and elliptic flow. We compared with recent STAR data. Future efforts will explore alternate vector spectral densities, will exploit recent advances in our knowledge of initial states, and will consider LHC conditions.

\section*{Acknowledgments}
We  thank I. Kozlov, J.-F. Paquet, L. Ruan , J. Zhao, and R. Vogt for helpful discussions. 
This work was supported in part by the Natural Sciences and Engineering Research Council of Canada, in part by the US National Science Foundation under grant no. PHY-0969394, and in part by by the A.-v.-Humboldt foundation.

\section*{References}
\bibliographystyle{h-physrev3.bst}
\bibliography{Proceedings-qm2012-v3.2}

\end{document}